# Energy Considerations for Lifting the Greenland Ice-Melt from the Earth's Gravitational Well

Mark A. Wessels, Ph.D.
Collin County Community College, Preston Ridge Campus
Frisco, Texas
October 4, 2017

**Abstract**
Climatologists have calculated that a complete melting of the Greenland Ice Sheet would cause global sea levels to rise by some 7.2 meters (23.6 feet). This article investigates the possibility of physically removing this excess water from the surface of the Earth by lifting the water into space. The theoretical minimum amount of work for this task is calculated, and is found to be equal to the amount of solar energy intercepted by the Earth in just 32 years.

## Rationale

The Greenland Ice Sheet represents one of the largest storehouses of fresh water on Earth, and is believed to be melting at a rate unequaled since the end of the last Ice Age[1,2,3]. The Sheet contains a volume of some 2.85 million cubic kilometers (694,000 cubic miles) of fresh water ice[4], having a mass of some **2.81 x 10$^{15}$ tons**. A complete melting of this ice would cause global sea levels to rise by 7.2 meters (23.6 feet)[4]. As some two-thirds of the world's major cities lie within low-elevation coastal zones[5], the rising sea would displace many millions of people all over the world. Some climate-change models see the majority of the Greenland ice melting over the next 2000 years[3], but this estimate is controversial, and is subject to revision. For this reason, a wide range of melt durations is examined. Melting of Antarctic ice is not considered here.

The continents lack the storage capacity to accommodate such a titanic volume. The storage capacity of all underground water reservoirs is estimated to be 30% of all *fresh water* at the surface of the Earth, but fresh water itself accounts for only 2.5% of *all* surface water[6]. Given these facts, the continents do not present a viable means of accommodating the ice-melt.

One bold possibility remains: that of *physically removing this excess water from the surface of the Earth by lifting and depositing the water into space, beyond Earth's gravitational well*. This report calculates the minimum theoretical amount of work required to do so. This result is then compared to the total amount of solar energy that is intercepted by the Earth in one year. The stunning result is that the amount of time needed to collect the required energy *is comparable to the amount of time over which the melting itself is expected to occur*.

## Energy Calculations

The gravitational potential energy between the (nearly) spherical Earth and a test mass located at the Earth's surface is given by[7]

$$U = -G \frac{M_E M_{test}}{R_E} \quad \text{(Eq. 1)}$$

whereby $M_E$ is the mass of the Earth (5.98 x $10^{24}$ kg), $M_{test}$ is the mass of the test mass, $R_E$ is the average radius of the Earth (6.38 x $10^6$ m), and G is the Universal Constant of Gravitation (6.6732 x $10^{-11}$ N·m² / kg²). This expression also gives the minimum amount of work that is needed to remove the test mass completely from Earth's gravitational well. Evaluating this expression for 1 metric-ton yields a value of **62.5 x $10^9$ Joules/ton**. The total amount of work required to lift the (fresh water) Greenland ice-melt is then 62.5 x $10^9$ J/ton multiplied by 2.81 x $10^{15}$ tons, or **1.75 x $10^{26}$ Joules**. This amount is some 318,000 times greater than the total amount of energy consumed by all of human civilization in the year 2011[8].

An interesting result is found when this figure is compared to the total amount of solar energy intercepted by the Earth each year. Every square meter, positioned outside the Earth's atmosphere at normal incidence to the Sun, intercepts 1361 Joules/sec (Watts) of power (the Solar Constant) [9]. When taken over the entire projected area of the Earth ($\pi R_E^2$, some 128 million square km), the total intercepted power is nearly 175 x $10^{15}$ Watts (175 petawatts). Multiplying by the number of seconds in one year (31.56 million) yields the total amount of solar *energy* intercepted by the Earth in one year, **5.52 x $10^{24}$ J**.

This result is just 3.15% of the 1.75 x $10^{26}$ J figure. Therefore, if *all* the Earth-intercepted solar power were applied to this task (and the process were perfectly efficient), the entire mass could be lifted in just **31.8 years**. This result implies that if human civilization could capture and process solar energy on a planetary scale (a true Type I civilization), this task is theoretically possible within a human time-scale.

## Intercept Area Fraction

The calculations above demonstrate a theoretical limit. But obviously, we cannot capture and direct all the Earth's intercepted solar power to this purpose. Neither will any realistic system operate near 100% efficiency. For these reasons, computations that consider a range of much longer time scales and realistic efficiencies are needed.

We wish to calculate the fraction of Earth's interception area needed to lift the ice-melt as a function of both (1) lift duration and (2) system efficiency. This is done as follows:

The *theoretical average power* (work per time) required to lift the ice-melt is the total work W (1.75 x $10^{26}$ J) divided by the lift duration T (in seconds), or

$$P = W/T \quad \text{(Eq. 2)}$$

The *actual power* is found by dividing this by the *system efficiency* ($\varepsilon$):

$$P_a = P/\varepsilon \quad \text{(Eq. 3)}$$

The required *fraction of Earth's interception area* ($f_E$) is then the actual power divided by 175 petawatts, or

$$f_E = P_a/175 \; Petawatts \quad \text{(Eq. 4)}$$

Figure 1 plots these results for a range of realistic time scales and system efficiencies.

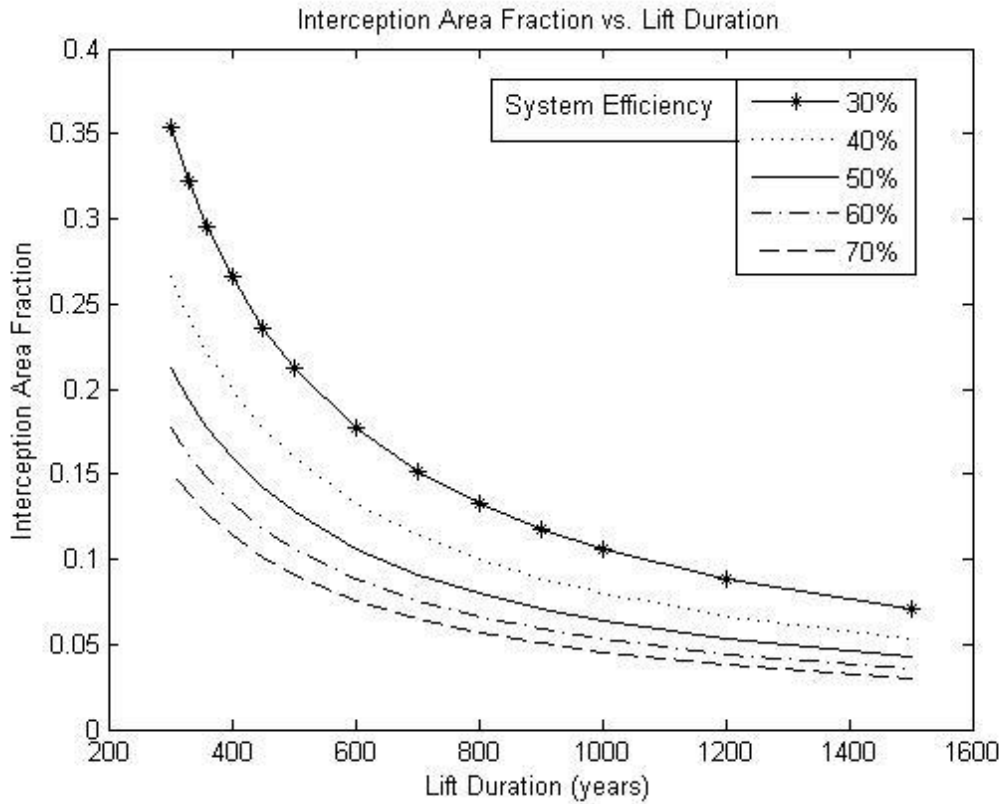

**Figure 1: Intercept Area Fraction versus Lift Duration**. This graph shows the fraction of the Earth's solar interception area required to capture the energy sufficient to lift the ice-melt mass beyond the Earth's gravitational well, as a function of both lift duration and system efficiency (shown in the legend). The family of curves follows a reciprocal-time (1/t) form. Better efficiencies and longer durations require smaller fractions. The values represent significant portions of the projected area of the Earth. The energy captured by tens of millions of square kilometers of solar collectors, working continuously for centuries, would be required.

## Rates of Mass Lift

The average rate of mass-lift is equal to the total mass ($2.81 \times 10^{15}$ tons) divided by the lift duration. These values are equally astonishing. See Figure 2 below.

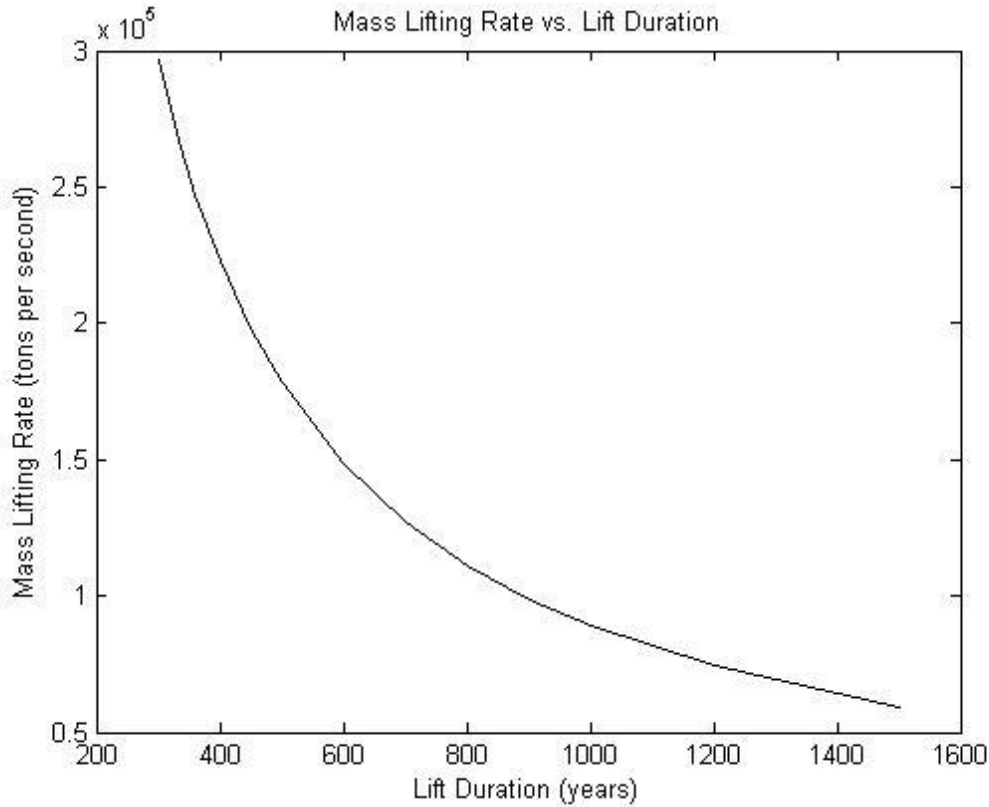

**Figure 2: Mass Lift Rates**. Distributing the lifting process over centuries still requires astonishing rates of mass lift – on the order of 125,000 *tons per second*. This calculation assumes a constant rate of lift over the indicated time. Conceiving of a machine having such capabilities challenges the human imagination.

## Implementation

Attempting a massive engineering project to capture a significant portion of the solar energy incident upon the Earth could significantly interfere with Earth's energy budget. For this reason, the collector would almost certainly need to be placed in space. A practical implementation of this concept would likely require a global arrangement of *space elevators*[10]. The elevator design would likely need to be that of a continuous loop, which revolves over a wheel located at the far end[11]. This loop could support a large number of water-filled containers, which would open and release their contents into space at the far end. (Rockets would be useless, as they are capable of launching only a small percentage of their initial mass into space.) The space elevator, however, remains an unrealized dream.

## Conclusion

The Earth intercepts a fantastic amount of power from the Sun – a fraction of which could be used to lift from the Earth's gravitational well the excess water expected from the melting of the Greenland Ice Sheet. The available power allows this to be achieved over the time scale of the melting itself. Thus, in principle, human civilization has a means to avoid the global flooding and horrible consequences that will result from the massive rise in sea level due to the melting. However, implementation of this concept will require the construction of space-elevator lifts, as well as the collection and management of solar energy on a planetary scale. Whether this is an achievable goal for humanity in the centuries to come remains to be seen.